\documentclass[letterpaper,aps,twocolumn,showpacs,superscriptaddress,
floats,prb]{revtex4}

\usepackage[english]{babel}
\usepackage[latin1]{inputenc}
\usepackage{amsmath, amssymb, amsfonts}
\usepackage{graphicx,epsfig,dsfont}
\usepackage{float}

\usepackage[squaren]{SIunits}
\usepackage{color}

\providecommand{\vect}[1]{\boldsymbol{#1}}
\providecommand{\V}{\boldsymbol{V_s}}

\begin{document}

\title{Rotating skyrmion lattices by spin torques and field or temperature gradients}

\author{Karin Everschor}
\affiliation{Institute of Theoretical Physics, University of Cologne, 
D-50937 Cologne, Germany
}
\author{Markus Garst}
\affiliation{Institute of Theoretical Physics, University of Cologne, 
D-50937 Cologne, Germany
}
\author{Benedikt Binz}
\affiliation{Institute of Theoretical Physics, University of Cologne, 
D-50937 Cologne, Germany
}
\author{Florian Jonietz}
\affiliation{Physik-Department E21, Technische Universit\"at M\"unchen,
D-85748 Garching, Germany}
\author{Sebastian M\"uhlbauer}
\affiliation{Forschungsneutronenquelle Heinz Maier Leibnitz (FRM II),  
Technische Universit\"at M\"unchen, D-85748 Garching, Germany}
\author{Christian Pfleiderer}
\affiliation{Physik-Department E21, Technische Universit\"at M\"unchen,
D-85748 Garching, Germany}
\author{Achim Rosch}
\affiliation{Institute of Theoretical Physics, University of Cologne, 
D-50937 Cologne, Germany
}

\begin{abstract}
Chiral magnets like MnSi form lattices of skyrmions, i.e. magnetic whirls, which
react sensitively to small electric currents $j$ above a critical current
density $j_{c}$. The interplay of these currents
with tiny gradients of either the magnetic field or the temperature 
can  induce a rotation of the magnetic pattern for $j>j_c$. 
Either a rotation by a finite angle of up to $\unit{15}{\degree}$ or -- for
larger gradients -- a continuous rotation with a finite angular velocity is
induced. We use Landau-Lifshitz-Gilbert equations extended by extra
damping terms in combination with a phenomenological treatment of pinning forces
to develop a theory of the relevant rotational torques. Experimental neutron
scattering data on the angular 
distribution of skyrmion lattices suggests that continuously rotating
domains are easy to obtain in the presence of remarkably small currents and temperature gradients.

\end{abstract}

\date{\today}

\maketitle

\section{Introduction: Spintorques and skyrmion lattices}
Manipulating magnetic structures by electric current is one
of the main topics in the field of spintronics. 
By strong current pulses one can, for example,  switch magnetic
domains in multilayer devices \cite{tsoi98,myer99},  induce microwave
oscillations in nanomagnets \cite{kise03} or
move ferromagnetic domain walls \cite{grol03,tsoi03}. The latter effect may be used
to develop new types of non-volatile memory devices \cite{Parkin:2008p8497}. It is therefore
a question of high interest to study the coupling mechanisms of currents to magnetic structures  \cite{slon96,berg96}.

Here, the recent discovery \cite{mueh09,Jonietz10} of the so-called skyrmion
lattice in chiral magnets like MnSi provides a new opportunity for studying the
manipulation of magnetism by electric currents both experimentally and
theoretically. The skyrmions in MnSi form a lattice of magnetic whirls, similar
to the superfluid whirls forming the vortex lattice in type-II superconductors.
While in ordinary ferromagnets, currents couple only to the canted spin
configurations at domain walls, the peculiar magnetic structure of the skyrmion
lattice allows for an efficient {\em bulk} coupling. Furthermore, the smooth
magnetic structure of the skyrmion lattice decouples efficiently from the
underlying atomic lattice and from impurities.
As a consequence, it was observed \cite{Jonietz10} that the critical current
density needed to affect the magnetic structure was more than five orders of
magnitude smaller than in typical spintorque experiments.  

These low current densities open opportunities for new types of experiments to
study quantitatively the physics of spin transfer torques. Due to the much lower
current densities it is now possible to perform spintorque experiments
in bulk materials and thus avoid the surface effects that dominate in nanoscopic
samples. Moreover for smaller currents the effects of heating and Oersted 
magnetic fields created by the current are suppressed.

\begin{figure}[t]
\begin{center}
\includegraphics[width=\linewidth]{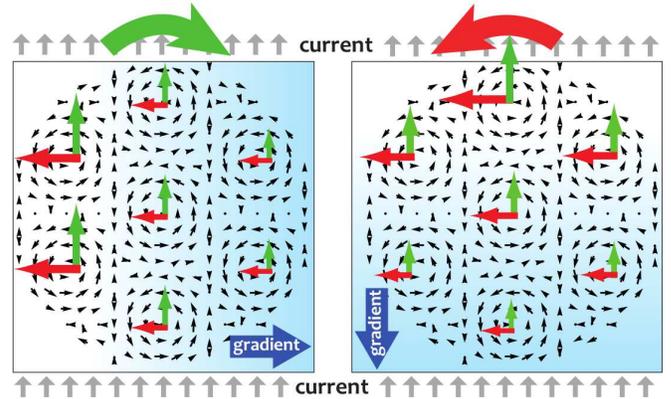}
\end{center}
\caption{Schematic plot of the forces on a skyrmion lattice perpendicular and
parallel
to the current flowing in vertical direction. For a static, non-moving skyrmion
lattice the  red horizontal arrows correspond to
the Magnus force and the green vertical arrows to dissipative forces.
In the presence of a temperature or field gradient, these forces change smoothly
across a domain, thereby inducing rotational torques which depend sensitively on
the relative orientation of current and gradient (and on the direction in which
the skyrmion lattice moves).  Small black arrows: local orientation of the
magnetization projected into the plane perpendicular to the magnetic field 
$\vect B$. In each unit cell the magnetization winds once around the unit sphere.  
\label{fig1}}
\end{figure}

In this paper we suggest experiments and develop a theory with the goal
to exploit the rotational motion instead of 
just translational motion to investigate the interplay of electric currents and
moving magnetic structures.   
Our theory is directly motivated by recent experiments \cite{Jonietz10},
where a change of orientation of the skyrmion lattice as a function of the 
applied electric current was observed with neutron scattering.
In Ref.~[\onlinecite{Jonietz10}] we have shown that the rotation arises from the interplay
of a tiny thermal gradient parallel to the current and the Magnus forces arising
from the spintorque coupling of current and skyrmion lattice. For example, the
rotation angle could be reversed by reversing either the current direction or
the direction of the thermal gradient.  

The basic idea underlying the theoretical analysis of our paper is sketched in
Fig.~\ref{fig1}.
In the presence of an electric current several forces act on the skyrmion
lattice. First, dissipative forces try to drag the
skyrmion lattice parallel to the (spin-) current. Second, the interplay of 
dissipationless spin-currents circulating around each skyrmion  and the
spin-currents induced by the electric current lead to a Magnus force  oriented
perpendicular to the current for a static skyrmion
lattice (for the realistic case of moving skyrmions the situation is more
complicated). In the presence of any gradient across the system (e.g. a
temperature
or field gradient), indicated by the color gradient, these forces will vary in
strength across a skyrmion domain.

 As in the experiment, we assume that the
gradients are tiny: on the length scale set by the skyrmion distance the
gradients have negligible effects. 
However, multiplying the tiny gradient with a large length, i.e., the size of a domain
of the
skyrmion lattice (which can be \cite{Adams2011} several hundred $\mu$m), one
obtains
a sizable variation of the forces across the domain. 
These inhomogeneous forces can give rise to rotational
torques. Whether the torque arises from the Magnus forces or the dissipative
forces depends, however, on the relative orientation of 
current and gradient and also on the direction in which the skyrmion lattice drifts. 
Fig.~\ref{fig1} gives a simple example: if, for example,
current and gradient are parallel to each other (right panel) the forces
perpendicular to the current direction (red horizontal arrows) give rise to
rotational torques while the parallel forces do not contribute.
The situation is reversed when current and gradient are perpendicular (left
panel).

We therefore suggest to use the rotation of magnetic structures as a function of
the relative orientation of current and further gradients as a tool to explore
the coupling of magnetism and currents. We will show that the resulting rotations depend very sensitively both on the relative size of the various forces affecting the skyrmion dynamics and on how these forces depend on the induced gradients. While we apply our theory here to
skyrmion lattices, our theoretical approaches can also be used for other complex
magnetic textures and our results should also have ramifications for other
setups \cite{Pribiag:2007p8556,Khvalkovskiy09}. Quantitatively, we will only
study the role of
gradients induced by changes in temperature or magnetic field but other options
are also possible. For example, macroscopic variations of the cross section of a
sample will lead to gradients in the current density. Also changes in the
chemical composition or strain in the sample can induce gradients.

It is also essential to investigate the effect of pinning of the magnetic
structure by inhomogeneities arising from crystalline imperfections.
Inhomogeneities distort the perfect skyrmion lattice and lead to forces
prohibiting (up to a very small creep)  the motion of the magnetic structure as
long as the current is below a critical value, $j<j_c$. Also for $j \gtrsim
j_c$,  inhomogeneities induce an effective, velocity dependent frictional force
on the moving skyrmion lattice connected to local, time-dependent distortions of
the skyrmion lattice. Pinning has widely been studied both experimentally and
theoretically for charge density waves and vortex lattices in superconductors
\cite{schm73, lark74, nattermann92,Blatter94}. As the dynamics of
skyrmions differs qualitatively (and quantitatively) from these two cases it is
not clear which of these results can be transferred to skyrmion lattices. 
Due to the non-linear dependence of the pinning forces on the velocity, they can
not be described by a simple damping term. Within this paper we will not try
to develop a theory of pinning but will instead use a simple phenomenological
ansatz to describe and discuss pinning effects.

 Rotational torques can also arise in the absence of the types of gradients
discussed above. In Ref.~[\onlinecite{everschor11}] we have studied
the role of distortions of the skyrmion lattice by the underlying atomic lattice
 extending the methods used by Thiele \cite{Thiele72} to rotational torques (this
method will also be used below). Such distortions indeed induce small rotational
torques in a macroscopically homogeneous system, i.e. without any external
gradients.  Similarly, also distortions induced by disorder can induce
rotational torques without external gradients as has been discussed in the
seminal paper by Hauger and Schmid \cite{schm73}. But all these effects
are very small and have {\em not} been observed in the experimental setup of
Ref.~[\onlinecite{Jonietz10}] as no rotation has been observed in the absence of
 gradients. Therefore they will be neglected in the following.

In the following we will first describe briefly the relevant Ginzburg-Landau
model and the Landau-Lifshitz-Gilbert equation
used to model the dynamics of the skyrmions. Here we include a novel damping
term $\alpha'$ recently introduced in Refs.~[\onlinecite{zhang09,nagaosa11}] (we
also add the corresponding $\beta'$ term). We then derive effective equations
for the translational and rotational mode where pinning physics is taken into
account by an extra phenomenological term.
This allows to develop predictions both for static rotations by a finite angle
and continuous rotations.  In the light of our results
we interpret experimental results on the angular distribution of skyrmion
lattices in the presence of currents and gradients.

\section{Setup}
\subsection{Ginzburg-Landau model}
 The starting point of our analysis is the standard  Ginzburg-Landau model 
of a chiral magnet in the presence of a Dzyaloshinskii-Moriya interaction \cite{naka80,bak80}.
After a rescaling of the length $\vect {r}$, the local magnetization 
$\vect {M} (\vect r)$ and the magnetic field $\vect B$ the free energy 
functional reduces to\cite{mueh09}
\begin{multline}
 F=\gamma_F \int\!d^3 r\,\left[ (1+t)\vect  M^2+ (\vect \nabla \vect M)^2 \right. \\
\left.  +2\,\vect  M\cdot(\vect \nabla\times \vect M)+\vect M^4 -\vect B\cdot \vect
M\right], 
\label{tF}
\end{multline}
Here $t \propto T-T_c^{\rm MF}$ parametrizes the distance to the mean-field phase transition at $B=0$
from a phase with helical magnetic order ($t<0$) to a paramagnetic phase ($t>0$)
\cite{naka80,bak80}. 
In the presence of weak disorder $t$ (and strictly speaking also the prefactors of
all other terms) fluctuates slightly as a function of $\vect r$.

The skyrmion lattice (stabilized by thermal fluctuations) exists
for a small temperature and field range\cite{mueh09}. It is translationally invariant
parallel to $\vect B$ and shows a characteristic winding of the magnetization in the plane 
perpendicular to $\vect B$, see Fig.~\ref{fig1}.

\subsection{Landau-Lifshitz-Gilbert equation}
To describe the dynamics of the orientation
$\hat{\vect \Omega}(\vect{r},t)=\vect{M}(\vect{r},t)/|\vect{M}(\vect{r},t)|$ of
the magnetization $\vect{M}(\vect{r},t)$ in the presence of spin-transfer
torques due to electric currents we use the standard Landau-Lifshitz-Gilbert
(LLG) equation,\cite{slon96,berg96,zhan04} extended by a  new dissipative term
\cite{zhang09,nagaosa11} 
\begin{multline} 
\label{LLG}
\left( \partial_t + {\vect v}_s \vect \nabla \right) {\hat{\vect \Omega}} 
= - {\hat{\vect \Omega}} \times {\vect H}_{\rm eff} 
+  \alpha \, \hat{\vect \Omega} \times \Bigl(\partial_t + \frac{\beta}{\alpha} 
{\vect v}_s \vect \nabla\Bigr){\hat{\vect \Omega}}\\
- \alpha'\Bigl[ \hat{\vect{\Omega}} \cdot \bigl( \partial_i \hat{\vect \Omega}
\times (\partial_t +\frac{\beta'}{\alpha'} \vect v_s \vect \nabla)\hat{\vect
\Omega}\bigr)
\Bigr]\partial_i \hat{\vect \Omega}.
\end{multline}
Here $\vect v_s$ is an effective spin velocity parallel to the spin current
density. More precisely, for smooth magnetic structures with constant amplitude
of the magnetization it is given by the 
ratio of the spin current \cite{footnoteSpinCurrent} 
and the size of the local magnetization, $|\vect{M}|$. In a good metal (for example, MnSi)
$\vect v_s$ is expected to be parallel to the applied electric current and to
depend only weakly on temperature and field.
The magnetization precesses in the effective magnetic field 
$\vect H_{\rm eff} \approx  -\frac{1}{ M} \, \frac{\delta F}{\delta
\hat{\vect \Omega}}$.  Strictly speaking
Eq.~(\ref{LLG}) is only valid for a constant amplitude of the magnetization,
$|\vect{M}|=\text{const}$. Since $|\vect{M}|$ varies only weakly\cite{mueh09}
in 
the skyrmion phase, we use as a further approximation  $\vect  H_{\rm eff}
\approx -\frac{1}{M} \frac{\delta F}{\delta \vect M}  \, \frac{\partial \vect
M}{ \partial
\hat{\vect \Omega}}$
where $M$  is the average local magnetization, $M^2=\langle \vect M^2 \rangle$.

The last two terms in Eq.~\eqref{LLG} describe dissipation.
$\alpha$ is called the Gilbert damping and
$\beta$ parametrizes the dissipative spin transfer torque.
The new damping term proportional to $\alpha'$ was introduced (for $\beta'=0$)
in Refs.~[\onlinecite{zhang09,nagaosa11}]. It arises from the ohmic damping of
electrons coupled by Berry phases to the spin texture as can be seen by rewriting 
Eq.~\eqref{LLG} in the form
\begin{multline}
 \label{LLG1}
-\frac{\delta F}{\delta
\hat{\vect \Omega}} = M \hat{\vect \Omega} \times 
\left( \partial_t + {\vect v}_s \vect \nabla \right) 
{\hat{\vect \Omega}}
+ {\alpha} M \Bigl(\partial_t +
\frac{\beta}{\alpha} {\vect v}_s \vect \nabla\Bigr){\hat{\vect \Omega}} \\
+M \hat{\vect \Omega } \times \alpha'\Bigl[\vect E^e_i + \frac{\beta'}{\alpha'}
\left(\vect v_s 
\times \vect B^e \right)_i \Bigr] \partial_i \hat{\vect \Omega}.
\end{multline}
where $\vect E^e_i= \hat{\vect \Omega} \cdot (\partial_i \hat{\vect \Omega} 
\times \partial_t \hat{\vect \Omega})$ can be interpreted as the emergent
electric 
field and $\vect B^e_i= \frac{1}{2} \epsilon_{ijk} \hat{\vect \Omega} \cdot 
(\partial_j \hat{\vect \Omega} \times \partial_k \hat{\vect \Omega})$
as the emergent magnetic field \cite{volovik,Schulz12}. These fields
describe the forces on the electrons arising from Berry phases which they pick up when
their spin adiabatically follows $\hat{\vect \Omega}(\vect r,t)$. They couple to
the spin rather to the charge:  electrons with magnetic moment parallel
(antiparallel) to $\hat{\vect \Omega}$ carry the ''emergent charge" $-1/2$ ($+1/2$),
respectively. For $\vect v_s=0$ the change of the free energy density
is given by
\begin{eqnarray}
\partial_t F=\frac{\delta F}{\delta
\hat{\vect \Omega}}\, \partial_t \hat{\vect \Omega} =- \alpha M (\partial_t \hat{\vect \Omega})^2
- \alpha' M (\vect E^e)^2.
\end{eqnarray}
which shows that the last term describes the dissipated power $\propto  (\vect E^e)^2$ arising from the emergent electric field.  $ \alpha' M$ is therefore approximately given by the spin-conductivity $\sigma_s$.

We have also added a new $\beta'$--term. The presence of such a term becomes evident if one considers the special case of a Galilean invariant system. In this case,  all forces have to cancel
when the magnetic structure is comoving with the conduction electrons,   $\hat{\vect \Omega}( \vect r,t)
=\hat{\vect \Omega}(\vect r - \vect v_s t)$. This is only possible for $\alpha=\beta$ and $\alpha'=\beta'$.
Solids are not Galilean invariant and therefore  $\beta'$ is different from  $\alpha'$ but one can, nevertheless, expect that the two quantities are of similar order of magnitude.

Which of the damping terms will dominate? 
As pointed out in Refs.~[\onlinecite{zhang09,nagaosa11}], the naive argument, that the $\alpha'$ terms are suppressed compared to the $\alpha$ terms as they contain two more derivatives, is not correct.
 The distance of skyrmions
is\cite{mueh09} proportional to $1/\lambda_{SO}$, where $\lambda_{SO}$
parametrizes the strength of spin-orbit coupling. 
While the $\alpha'$ term has two more gradients compared to the $\alpha$ term,
the contribution arising from $\alpha'$ is, nevertheless, of the same order in powers of $\lambda_{SO}$, if we assume that $\alpha$
arises only from spin-orbit coupling, $\alpha \propto \lambda_{\rm SO}^2$, while  $\alpha' \propto  \lambda_{\rm SO}^0$ (ohmic damping (see above) does not require spin-orbit effects). As furthermore
$\alpha$ is proportional to a scattering rate while $\alpha'$ is proportional to a conductivity and therefore
the scattering time\cite{zhang09,nagaosa11}, $\alpha'$ and $\beta'$ might be  the dominating damping
terms in good metals. 

\section{Dynamics of skyrmions}
 
Our goal is to describe both the drift and the rotation of the skyrmion lattice
in the limit of small current densities and small magnetic or thermal
gradients. 
We therefore assume that $\vect v_s$ is small compared to all characteristic
velocity scales of the skyrmion lattice (e.g. $T_c-T$ multiplied with the
skyrmion distance). 
The gradients should be so small that the total change across a domain of radius
$r_d$ remains small, $r_d \vect \nabla \lambda \ll \lambda$ where $\lambda$ is $B$ or
$T_c-T$ for magnetic or thermal gradients, respectively. In this limit, both the
drift velocities $ v_d \lesssim v_s$ and the angular velocity $\partial_t \phi
\propto \vect v_s \cdot  \vect \nabla \lambda$ characterizing rotational motion remain 
small. Below we will show, that even $\partial_t \phi \, r_d$, the velocity at 
the boundary of the domain remains small in the considered limit.

We  can therefore neglect macroscopic deformations of the magnetic structure and
 consider  the following ansatz
\begin{eqnarray}
\hat{\vect \Omega}(\vect r,t)=\vect R_{\phi (t)}\cdot  \hat{\vect \Omega}_0\!
\left(\vect R_{\phi(t)}^{-1} \cdot (\vect r- \vect v_d t)\right)
\end{eqnarray}
Here  $\hat{\vect \Omega}_0(\vect r)$ describes the static skyrmion lattice, 
$\vect R_{\phi}$ is a matrix describing a rotation by the angle $\phi$
around the direction of the skyrmion lines (i.e. around the field direction when
anisotropies are neglected, which will be assumed in the following) and $\vect v_d t$ describes the location of the center of the skyrmion domain. This ansatz
describes  a magnetic domain which
rotates around its center, while the center is moving with the velocity $\vect
v_d$. When the torque forces are too weak to induce a steady-state rotation,
such that $\partial_t \phi=0$, we will study rotations by the finite angle $\phi$ as in
the experiment of Ref.~[\onlinecite{mueh09}].

\subsection{Drift of domains}

To obtain an equation for the drift velocity $\vect v_d$  we
follow Thiele \cite{Thiele72} and project Eq.~(\ref{LLG1}) onto the translational
mode
by multiplying Eq.~(\ref{LLG1})
with $\partial_i \hat{\vect \Omega}$ and integrating over a unit cell (UC). We
thereby obtain to order $(\vect \nabla \lambda)^0$ (where no rotations occur) an equation for
the force per 2d magnetic unit cell (and per length) \cite{footnote1}
\begin{align} 
\vect G &\times ( {\vect v}_s-{\vect v}_d ) 
+ \vect{\mathcal D} (\tilde{\beta}{\vect v}_s-\tilde{\alpha} {\vect v}_d )
+\vect F_{\mathrm{pin}} =0
 \label{drift1} \\
\vect{G}_i &= \int_{\rm UC} d^2 r\, M \vect B^e_i 
=\mathcal G \hat{\vect {B}}_i, \quad \mathcal G= 4 \pi M W \nonumber 
\\
{\vect {\mathcal{ D}}}_{ij} &= \int_{\rm UC} d^2 r\, M \partial_i
{\hat{\vect\Omega}}\,
\partial_j {\hat{\vect \Omega}} = \mathcal D \vect P_{ij} \nonumber \\
\mathcal{ D}'&=  \int_{\rm UC} d^2 r\, M\, (\vect B^e)^2  \nonumber \\
\tilde \alpha&= \alpha + \alpha' \mathcal{D'}/\mathcal{D}  
\quad \text{and} \quad \tilde \beta = \beta + \beta' \mathcal{D'}/\mathcal{D} \nonumber
\end{align}

Here the first term describes the Magnus force which is proportional to the topological winding number $W$
which is for the skyrmion lattice exactly given by $W=-1$. $\vect G$ is called
the gyromagnetic coupling vector following Thiele \cite{Thiele72}. The second term 
are the dissipative forces with the projector  $\vect P$ into the plane
perpendicular to $\bf B$,  $\vect P= (\mathds{1} - \hat{\vect B}\cdot
\hat{\vect{B}}^T)$. 

Besides the forces discussed above, also pinning forces, described by the last
term in  Eq.~\eqref{drift1}, have to be considered. Formally, they are encoded in
spatial fluctuations of $\delta F/\delta \hat{ \vect \Omega}$ in Eq.~\eqref{LLG1}. The Thiele
approach, used above, which considers only a global shift (or a global
rotation\cite{everschor11}, see below) of the magnetic structure does not
capture these pinning effects as for a perfectly rigid magnetic structure,
random pinning forces average to zero, such that no net effect remains in the
limit of a large domain. To describe pinning, it is necessary
\cite{schm73,lark74} to take into account that the magnetic
structure adjusts locally to the pinning forces, a complicated problem for which
presently no full solution exists\cite{Blatter94,Kopnin02} and which is far beyond the
scope of the present paper. Instead, we use a phenomenological ansatz and write
for a finite drift velocity $\vect v_d$
\begin{eqnarray}\label{pinningF}
\vect F_{\mathrm{pin}}= -4 \pi M v_{\mathrm{pin}} f(v_d / v_{\mathrm{pin}})\,
\hat{\vect v}_d
\end{eqnarray}
to describe a net pinning force, which is oriented opposite to the direction of
motion. Its strength, which depends both on the number (and nature) of defects responsible for pinnning 
and the elastic properties of the skyrmion lattice, is parametrized by the `pinning velocity'
$v_{\rm pin}$. The function $f(x)$ with $f(x\to 0)=1$ and $f(x\to \infty)=x^\nu$
parametrizes the non-linear dependence of the pinning force on the velocity.
Presently, it is not clear to what extent $f(x)$ depends on microscopic details
and also the exponent $\nu$ is not known. For large driving velocities, however,
pinning becomes less and less important ($\nu<1$)\cite{Blatter94,Kopnin02,nattermann92}. 
If the driving forces are smaller than the force $4 \pi M v_{\rm pin}$, needed to depin
the lattice, $v_d$ vanishes and the pinning forces cancel exactly the driving
forces. Note that we do not consider creep, i.e. a tiny motion driven by thermal
(or quantum) fluctuations, which occurs even in the pinning regime
\cite{Blatter94}.
If the dissipative forces can be neglected, it is in principle possible to
obtain $f(x)$ from a measurement of the velocity of the skyrmion
lattice\cite{Schulz12}.

In the limit $v_s \gg v_{\rm pin}$, where $\vect F_{\mathrm{pin}}$ can be
neglected, we solve Eq.~\eqref{drift1} for $\vect v_s \perp \vect B$ to obtain
\begin{equation}
\label{vd}
{ \vect  v_d}=
 \frac{\tilde \beta}{\tilde \alpha} 
  { \vect v_s}+ 
\frac{\tilde \alpha-\tilde \beta}
 {\tilde  \alpha^3 (\mathcal D/\mathcal G)^2+\tilde  \alpha} 
 \Bigl( { \vect v_s} + \tilde  \alpha  \frac{\mathcal D}{\mathcal G}
 \hat{\vect{B}} \times {\vect v_s}\Bigr) 
\end{equation}
with $\tilde \alpha= \alpha + \alpha' \mathcal{D'}/\mathcal{D}$ and
$\tilde \beta = \beta + \beta' \mathcal{D'}/\mathcal{D}$ .

\subsection{Rotational torques}
By symmetry, a small uniform current cannot induce any
rotational torques on a skyrmion lattice with perfect sixfold rotation symmetry
and therefore all effects arise from gradients.
To derive an equation for the rotational torques which determine the rotations
around the $\vect B$ axis, we follow\cite{everschor11} a similar procedure as
used for
the translations by multiplying (\ref{LLG1}) by the generator of rotations applied to $\hat{\vect \Omega}$
\begin{eqnarray}
\partial_\phi
\hat{\vect\Omega} = \hat{\vect{B}} \times
\hat{\vect \Omega} - (\hat{\vect B} ({\Delta \vect r} \times \vect \nabla) )
\hat{\vect \Omega} \label{rotp}
\end{eqnarray}
 with $\Delta \vect r=\vect r -\vect v_d t$ and integrating over $\vect r$. 
This procedure leads to several types of contributions.

For the first type of contribution, we observe that the second term in
Eq.~(\ref{rotp}),
linear in $\Delta \vect r$, is much larger than the first one which we can
therefore neglect whenever the second term contributes. The second term induces
torques of the form $\vect r \times \vect f$ where the force $f_i$ is obtained
by multiplying $\nabla_i
\hat{\vect \Omega}$ with the terms of Eq.~(\ref{LLG1}). In the presence of
gradients of the parameter $\lambda$ we obtain
\begin{align}
\int \hat{\vect B}\cdot [\vect r \times \vect f(\lambda(\vect r))]  &\approx \int 
\left(\hat{\vect B} \cdot [\vect r \times \partial_\lambda \vect f ]\right) \left(\vect r \cdot \vect 
\nabla \lambda\right) \nonumber \\
&\approx \frac{A}{4 \pi}  \hat{\vect B}\cdot [\vect \nabla \lambda \times  
\partial_\lambda  \int  \vect f  ] \label{expand}
\end{align}
where  $A$ is the area of the domain. Here it is essential to take the derivative
with respect to $\lambda$ for {\em fixed} $\vect v_d$ reflecting that due to the
rigidity of the skyrmion crystal $\vect v_d$ is constant across the domain. As
the sum of all relevant forces vanishes [Eq.~\eqref{drift1}],
$\sum_i \vect f_i(\lambda,\vect v_d)=0$, one obtains $\frac{d}{d \lambda} \sum_i
\vect f_i=0$ while $\left. \frac{\partial}{\partial \lambda} \sum_i \vect
f_i\right|_{\vect v_d}$ is finite. In Eq.~\eqref{expand} we have implicitely
assumed a symmetrically shaped domain, where integrals odd in $\vect r$ vanish.
In general, there will also be a 
shape dependent torque $\mathcal T_{\mathrm{shape}}$ arising even in the absence
of a gradient. As its sign is random, it can easily be distinguished from the
other torques (and appears to be relatively small in the MnSi experiments
\cite{mueh09}). More difficult is the question what happens at the interface of
different domains or when a domain comes close to the surface of the sample.
Nominally surface forces are suppressed by a factor proportional to $1/\sqrt{A}$
compared to the bulk terms considered above but the relevant prefactors are
difficult to estimate.
We will neglect in the following formulas both extra surface forces and shape
dependent torques.

A different contribution arises from the time derivatives $\partial_t \hat{\vect
\Omega}=
\partial_t \phi \,
\partial_\phi \hat{\vect \Omega}-(\vect v_d  
\vect \nabla) \hat{\vect \Omega}$ in Eq.~(\ref{LLG1}). The contribution proportional to $\vect v_d$ is of the form discussed above.
The term proportional to $\partial_t \phi$ leads to extra torques independent of $\vect \nabla \lambda$.
By combining the linear term in $\Delta r$ from $\partial_\phi \hat{\vect \Omega}$ with the second term of Eq. (\ref{rotp})
we obtain for example the contribution $\alpha \, \partial_t \phi \int M [(\hat{\vect B} [\Delta \vect r
\times  \vect \nabla])\vect \Omega ]^2$ which is also linear in $A$. Physically this term describes the frictional torque
which is linear in the angular
velocity $\partial_t \phi$. The frictional torque per volume is proportional  to $A$ because the velocity and therefore the
frictional forces grow linearly with the distance from the center of the rotating domain.

Finally, a contribution exists which is independent of the
gradients $\vect \nabla \lambda$, the angular velocity $\partial_t \phi$ and  of
$\vect v_s$. This contribution describes that in the absence of any external
perturbation the skyrmion lattice has a preferred orientation relative to the
atomic lattice. Such terms express that
angular momentum can  be transferred directly from the skyrmion lattice to the
underlying atomic lattice mediated by spin-orbit coupling and small anisotropy
terms (not included in Eq.~\eqref{tF}). These terms have been discussed in detail in Ref.~[\onlinecite{everschor11}].
This torque per unit cell
\begin{equation}
\label{LatticeTorque}
\mathcal{T}_L = -  \int_{\rm UC} d^2 r\, \frac{\delta F}{\delta \hat{\vect \Omega}}
(\hat{\vect{G}}_{\rm rot} {\hat{\vect \Omega}})
=-\frac{\partial F_{\rm UC}}{\partial \phi}\approx -\chi\, \sin(6 \phi)
\end{equation}
can be expressed by the change of free energy per unit cell, $F_{\rm UC}$,
upon rotation by the angle $\phi$, where $\phi=0$ reflects the equilibrium
position and $\sin 6 \phi$ reflects the sixfold symmetry of the skyrmion
lattice. 
  As has been discussed in Ref.~[\onlinecite{Jonietz10}], the absolute value of
$\chi$ in materials like MnSi is tiny as it arises only to high order in
spin-orbit coupling and, in contrast to all other terms, it is not linear in the
size of the domain. Nevertheless, we have to consider this term, as it is the
leading contribution arising to zeroth order in $\vect \nabla \lambda$ and $\vect
v_s$.

Balancing all torques (per unit cell) we obtain as our central result
\begin{align}\label{TTT}
0 &=\mathcal T_L +\mathcal T_G + \mathcal T_{\mathrm{pin}} + \mathcal T_{\mathcal
D}\\
\mathcal T_G&= \frac{A}{4 \pi} \vect \nabla \lambda
\cdot \left[ \frac{ \partial( \mathcal G \vect v_s)}{\partial \lambda}- \frac{
\partial \mathcal G }{\partial \lambda} \vect v_d\right]\nonumber \\
\mathcal T_{\mathrm{pin}}&=\frac{A}{4 \pi}  
 \vect \nabla
\lambda \cdot [ \hat{ \vect{B}} \times \hat{\vect v}_d]\frac{ \partial F_{\rm
pin}}{\partial \lambda}, \quad F_{\rm pin} \equiv |\vect F_{\rm pin}| \, \nonumber \\
\mathcal T_{\mathcal D}&=-\frac{A  \tilde{\alpha}  \mathcal D }{2 \pi} 
\partial_t \phi \nonumber \\
& \quad - \frac{A }{4 \pi}   \vect \nabla \lambda \cdot 
\biggl[\hat{ \vect{B}} \times \Bigl(  \frac{ \partial (\mathcal D \tilde{\beta}
\vect  v_s)}{\partial \lambda} 
-  \frac{ \partial (\mathcal D \tilde{\alpha})}{\partial \lambda}  \vect v_d
\Bigr)\biggr] \nonumber 
\end{align}
The direction of the torques, which depends on the relative orientation of
velocities and currents, is for $\vect v_d=0$ (and $\partial_t \phi=0$) fully consistent 
with the simple picture shown in Fig.~\ref{fig1}:
the dissipative torques $\mathcal T_{D}$ arise when
gradient and current are perpendicular to each other while
the reactive torque $\mathcal T_G$ arising from the Magnus force is activated
for a parallel alignment of gradients and currents. For finite $\vect v_d$, however, this simple intuitive picture cannot be used especially as some of the torques tend to cancel when $\vect v_d$ approaches $\vect v_s$.

\subsection{Rotation angle and angular velocity}
\label{phiandomega}

\begin{figure}[t]
\begin{center}
\includegraphics[width=\linewidth]{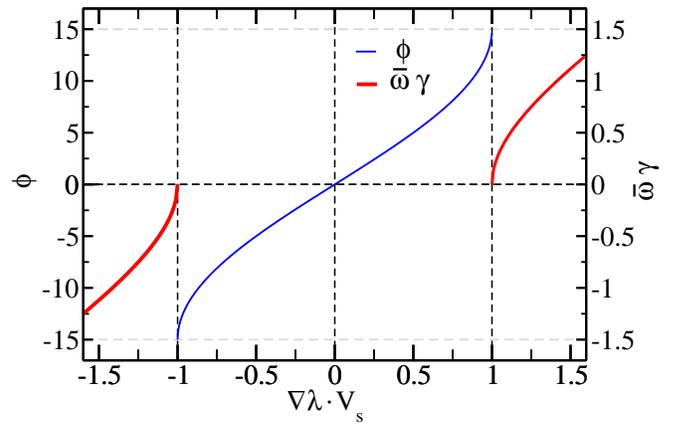}
\end{center}
\caption{Rotation angle $\phi$ (in units of $\unit{1}{\degree}$) and angular velocity $\bar \omega$ (times the prefactor $\gamma$)
as a function of $\vect \nabla \lambda \cdot \V$ determined from Eq.~\eqref{compactTTT}.}  
\label{figrot}
\end{figure}

Eq.~\eqref{TTT} can be rewritten in the compact form
\begin{eqnarray}
\label{compactTTT}
\sin 6 \phi =- \gamma \,\partial_t \phi + \vect \nabla \lambda \cdot \V
\end{eqnarray}
where  $\gamma=\frac{A  \tilde{\alpha} \mathcal D }{2 \pi \chi}$ and the vector 
$\V=\V[\vect v_s]$  can be obtained by first solving Eq.~\eqref{drift1} to obtain
$\vect v_d$ as a function of $\vect v_s$. This function is inserted into Eq.~\eqref{TTT} which, finally, is devided by $-\chi$.
The function $\V[\vect v_s]$ with $\V[0]=0$ is proportional to the area $A$ of the domain and encodes all information how the current couples 
to small gradients and includes contributions from Magnus forces, dissipative forces and pinning.

\subsubsection{Dependence on size of gradients}

\begin{figure}[t]
\begin{center}
\includegraphics[width=\linewidth]{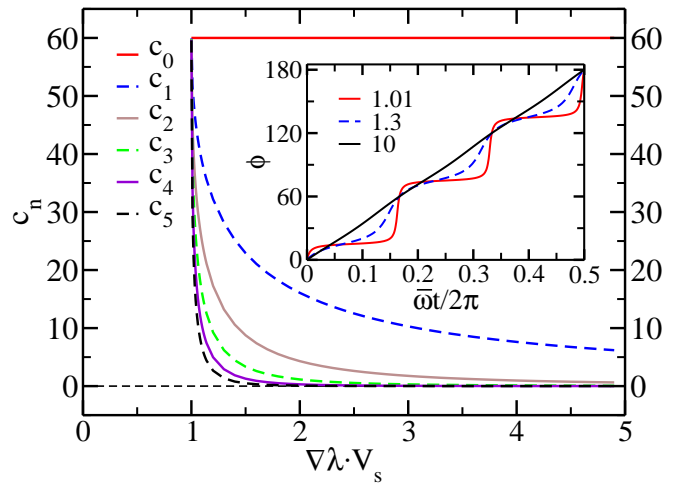}
\end{center}
\caption{Inset: Rotation angle (in units of $\unit{1}{\degree}$) as a function of time for three values of $\vect \nabla \lambda  \, \vect \V>1$, see Eq.~(\ref{phit}). For torques close to the value where rotations sets in, the rotation is strongly anharmonic. This
can also be seen by considering the Fourier coefficients  
$c_n=|\int_0^{2 \pi/ 6 \bar \omega} \partial_t \phi e^{i n 6 \bar \omega t} \, dt|$ shown in the main panel as a function of $\vect \nabla \lambda  \, \vect \V$.
\label{figAngle}}
\end{figure}

Qualitatively, three different regimes 
have to be distinguished. For $j<j_c$,
when pinning forces cancel all reactive and dissipative forces, there is neither
a motion nor a rotation of the skyrmion lattice, $\V=0, \phi=0$, within our approximation. Note, however, that it is well known from the physics of charge density waves or vortices \cite{Blatter94} that even
below $j_c$ a slow creep motion is possible. Whether during this creep also rotations are possible
is unclear, but the rather sharp onset of the rotation in the experiments of Ref.~\onlinecite{Jonietz10}, see
Fig.~\ref{figExp}, seems to contradict a scenario of pronounced rotations during creep. 
For $j>j_c$, the domains move and $\V$ will generally be finite. In this case,
one can control the size and direction of rotations by the  size of $\vect \nabla
\lambda$ as shown in Fig.~\ref{figrot}. 
For  $| \vect \nabla \lambda
\cdot \V |<1$, one obtains
a solution where $\partial_t \phi=0$ but the gradients induce
a rotation by a finite angle
\begin{eqnarray}
\phi=\frac{1}{6} \arcsin \vect \nabla \lambda
\cdot \V, \label{phi}
\end{eqnarray} which grows upon increasing $\vect \nabla \lambda$ from zero until it reaches 
the maximal possible value  $\pi/12=15^\circ$  (rotations by an
average angle of 10$^\circ$ have
already been observed \cite{Jonietz10}, see Fig.~\ref{figExp}). For $| \vect \nabla \lambda
\cdot \V |>1$
the domain rotates (see Fig.~\ref{figrot}) with the (average) angular velocity 
\begin{eqnarray}\label{omega}
\bar \omega=\frac{\sqrt{(\vect \nabla \lambda \cdot \V)^2-1}}{\gamma}
\end{eqnarray}
and
Eq.~(\ref{compactTTT}) is solved by
\begin{eqnarray}\label{phit}
\phi(t)=\frac{1}{3} \arctan\!\left[ \frac{1+\gamma \bar \omega \tan( 3 \bar \omega t)}{\sqrt{1+\gamma^2 {\bar \omega}^2}}\right].
\end{eqnarray}
displayed in the inset of Fig.~\ref{figAngle}. As both $\gamma$ and $\V$ are linear in the area $A$ of the domain, $\bar \omega \approx (\vect \nabla \lambda \cdot \V)/\gamma$ becomes independent of
the domain size for $A\to \infty$.  In this limit, the domain rotates continuously, $\phi=\bar \omega t$.
Close to the threshold, $\vect \nabla \lambda \cdot \V=1$, however, the rotation becomes very slow close to an angle of $\unit{15}{\degree}$ (plus multiples of  $\unit{60}{\degree}$).

A way to detect the rotation of the magnetization is to exploit the emergent
electric field $\vect E^e$ which obtains a contribution proportional to 
$\propto \partial_t \phi$ and can be measured in a Hall experiment
\cite{Schulz12}. In Fig.~\ref{figAngle} we therefore show the modulus of the
Fourier components,
$|c_n|=|\int e^{i 6 \bar \omega n t} \partial_t \phi \, d t|$ of $\partial_t
\phi$
as a function of $\vect \nabla \lambda
\cdot \V$. At the threshold, all Fourier components are of equal weight while
for 
large gradients the rotation gets more uniform.

For fixed $\bar \omega$ the velocities at the boundary of the domain, $v_{\rm
b}=\bar \omega r_d$, grow linearly
with the radius of the domain $r_d$. As we have assumed that the gradients
across the sample and therefore also across a single domain are small, $ r_d\,
\vect \nabla \lambda \ll \lambda$, the velocities nevertheless remain small,
$v_{\rm b}\ll |\V| \lambda/\gamma \lesssim v_s/\tilde \alpha$. While our
estimate does not rule out that $v_{\rm b}$ can become somewhat larger than
$v_s$ or $v_d$, we expect that the typical situation is that the velocity
$v_{\rm b}$ arising from the  rotation remains smaller than the overall drift
velocity of the domain $v_d$. This estimate also implies that violent phenomena
like the breakup of domains due to the rotation will probably not occur.

\subsubsection{Domain size dependence and angular distribution}

In a real system, there will always be a distribution of domain sizes $A$. Both
$\V$ and $\gamma$ are linear in $A$ and therefore both the rotation angle
(\ref{phi}) and the angular velocity (\ref{omega}) will in general depend on the
domain size and therefore on the distribution of domains. 

Only in the limit $| \vect \nabla \lambda \cdot \V | \gg 1$, the dependence on
$A$ cancels in Eq. (\ref{omega}) and all domains rotate approximately with the
same angular velocity.
For  $| \vect \nabla \lambda \cdot \V | \lesssim 1$ one will in general obtain a
distribution of rotation angles which can be calculated from the distribution of
domain sizes $P_d(A)$. For the static domains only angles up to $\unit{15}{\degree}$ are possible with 
\begin{align}
 P^s_\phi &= \int_0^{A_c} d A\,  P_{d}(A) \delta ( \phi -
\frac{\arcsin(A/A_c)}{6}) \nonumber\\
&= 6 A_c \cos(6 \phi) \, P_{d}(A_c \sin(6 \phi)) \quad {\rm for}\ 0\le \phi \le \frac{\pi}{12} \label{Pphi}
\end{align}
where $A_c=A/(\vect \nabla  \lambda \cdot\V)$ is the size of a `critical' domain which
just starts to rotate continuously.

\begin{figure}[t]
\begin{center}
\includegraphics[width=\linewidth]{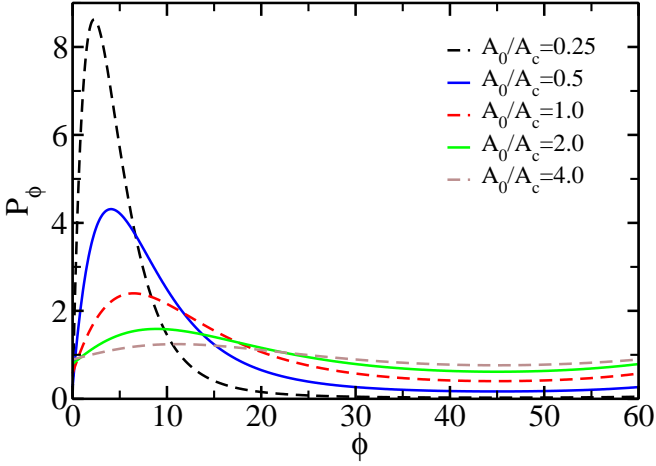}
\end{center}
\caption{Angular distribution $P_\phi$ of the rotation angle of the skyrmion
lattice for various values of $A_0/A_c \propto \nabla \lambda$ (see text). Here
we assumed a distribution of domain sizes of the form $P_{d}(A)=e^{-A/A_0}\frac{
A}{A_0^2}$. While static domains contribute only for $0\le \phi \le
\unit{15}{\degree}$ one obtains a smooth angular distribution when one takes the
rotating domains with $A>A_c$ into account.
\label{figDomains}}
\end{figure}

The continuously rotating domains also have a non-trivial angular distribution
as their rotation will be slowed down when the counterforces are strongest,
i.e., for $\phi=\unit{15}{\degree}$, see inset of Fig.~\ref{figAngle}. 
The angular distribution, $P^r_\phi$, of the rotating domains
is calculated from
distribution of domain sizes, $P_d(A)$, and the angular distribution, $p^r_\phi(A)$,
of a single domain
\begin{eqnarray}
P^r_\phi&=&  \int_{A_c}^\infty d A\,  P_{d}(A) \,p^r_\phi(A) \nonumber\\
p^r_\phi(A)&=&\frac{1}{T} \int_0^T \delta(\phi-\phi(t)) dt=\left.\frac{1}{T
\partial_t \phi}\right|_{\phi(t)=\phi} \nonumber \\
&=&\frac{3}{\pi} \frac{\sqrt{A^2-A_c^2}}{ A-A_c \sin 6 \phi}
\end{eqnarray}
where $T=2 \pi/(6 \bar \omega)$.
While both $P^s_\phi$ and $P^r_\phi$ are non-analytic at 
$\phi=\unit{15}{\degree}$, the total distribution, $P_\phi=P^s_\phi+P^r_\phi$ is
smooth for $\phi>0$ and normalized to 1, $\int_0^{2 \pi/6} P_\phi\, d\phi=1$.
 In Fig.~\ref{figDomains} we show $P_\phi$ assuming the domain distribution
$P_{d}(A)=e^{-A/A_0}\frac{ A}{A_0^2}$ for various values of $A_0/A_c$.

In elastic neutron scattering, the  skyrmion phase is observed by six Bragg
spots forming a regular hexagon in a plane perpendicular to the magnetic field.
A rotation of the skyrmion domain results in a rotation of these Bragg spots. 
Therefore the angular distribution $P_\phi$ of rotation angles is directly
observable (see Sec.~\ref{experiments} below) by measuring the scattering intensity as a function of angle. By
comparing angular distributions for different strength of the current or
gradient, one can -- at least in principle -- obtain not only $A_c$ as a
function of $\vect \nabla \lambda$ or $j$ but also the distribution of domain
sizes. 
 The latter can be extracted most easily in the regime where most of the domains
do not rotate continuously
 by plotting $P_\phi/\cos 6 \phi$ as a function of $\sin 6 \phi$ using
Eq.~(\ref{Pphi}).

\subsubsection{Dependence on strength of current}
While the behavior of $\phi$ and $\bar \omega$ as a function of $\vect \nabla
\lambda$ is rather universal and independent
of microscopic details, its dependence on the strength of the current for fixed
$\vect \nabla \lambda$ is much more complex.
As discussed above, $ \V=0$ for $j<j_c$. Directly at $j_c$, when the domain
starts to move with $v_d \approx 0$, $\V$ jumps to the finite value
\begin{multline}
\V|_{v_s= v_{\mathrm{pin}}} =- \frac{A}{4 \pi \chi} 
\left[ \left(-\frac{\partial \mathcal{G} \vect{v}_s }{\partial \lambda} 
+ \frac{\mathcal{G} \vect{v}_s }{F_{\mathrm{pin}}} \frac{\partial F_{\mathrm{pin}}}{\partial \lambda}\right)\right. \\
\left. + \hat{\vect B} \times 
\left( \frac{\partial \mathcal{D} \tilde{\beta} \vect{v}_s}{\partial \lambda} 
-\frac{\mathcal{D} \tilde{\beta} \vect{v}_s }{F_{\mathrm{pin}}} 
\frac{ \partial F_{\mathrm{pin}}}{\partial \lambda} 
\right)\right].
\end{multline}
Note that the jump is independent of $\alpha$ and $\alpha'$ as well as of their gradients, as the 
skyrmions are not moving directly at the depinning transition (see Fig.~\ref{figRotJTheory2}).
Depending on the direction and size of $\vect \nabla \lambda$, the jump of $\V$ either leads
to a jump of the rotation angle for 
$| \vect \nabla \lambda \cdot \V |<1$ or immediately to a continuous rotation
for $| \vect \nabla \lambda \cdot \V |>1$.

Upon increasing the current, $ \vect \nabla \lambda \cdot \V $ can either
increase, decrease or even change its sign depending on (i) the direction of
$\vect \nabla \lambda$ and (ii) on the question which of the forces changes most strongly
when varying $\lambda$ (i.e., temperature or magnetic field).

\begin{figure}[t]
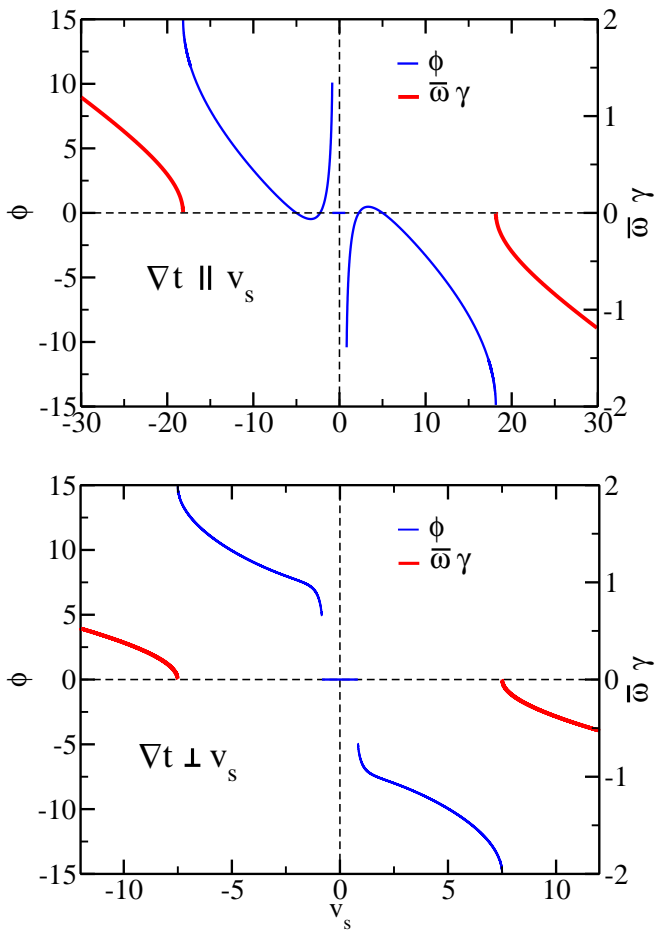

\begin{center}
\includegraphics[width=  \linewidth]{figure5}\\
\vspace{1mm}
 \includegraphics[width=  \linewidth]{figure6}
\end{center}
\caption{Rotation angle $\phi$ (in units of $\unit{1}{\degree}$) and angular velocity, $\gamma \bar \omega$, as a
function of $v_s$ for a temperature gradient {parallel} ($\vect \nabla t \|
\vect v_s$, $\vect \nabla t=(-0.1,0,0)$, upper panel) and perpendicular ($\vect \nabla
t \perp \vect v_s$,  $\vect \nabla t=(0,-0.05,0)$, lower panel) to the current 
($\alpha=0.2, \beta=0.45, \alpha'=0.01, \beta'=0.2, A/\chi=200, t=-1, \vect
B=(0,0,1/\sqrt{2}), v_{\rm pin}=1, f=1$).
For both geometries one observes a  jump of $\phi$ at $v_s \approx v_{\rm pin}$
from zero to a finite rotation angle. After the initial jump the rotation angle
increases for the perpendicular configuration (panel b) while for the parallel
arrangement first a drop and then an increase up to the maximal angle of
$\unit{15}{\degree}$ occurs. For larger $v_s$ a continuous rotation
characterized by the angular velocity $\bar \omega$ sets in for both configurations.
For the calculation we assumed that the damping parameters and $v_{\rm pin}$ are
independent of $t$. 
\label{figRotJTheory1}}
\end{figure}

\begin{figure}[t]
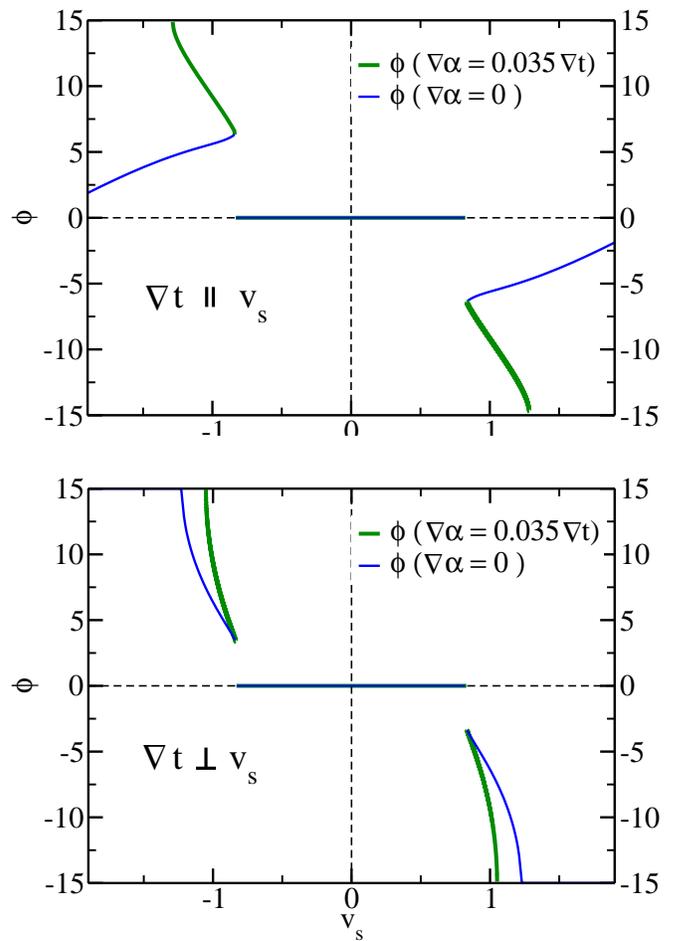

\begin{center}
\includegraphics[width= \linewidth]{figure7}\\
\includegraphics[width=  \linewidth]{figure8}
\end{center}
\caption{Rotation angle $\phi$ (in units of $\unit{1}{\degree}$) as a function of 
$v_s$ for a temperature gradient
parallel ($\vect \nabla t \| \vect v_s$, upper panel) and perpendicular ($\vect
\nabla t \perp \vect v_s$,  lower panel) to the current. The parameters are the
same as in Fig.~\ref{figRotJTheory1}
with two exceptions. First, we have taken into account that in the experiments
of Ref.~\onlinecite{Jonietz10} the temperature gradient grow with the square of
the applied current,
$\vect \nabla t = (-0.1 v_s^2, 0, 0)$ and $\vect \nabla t = (0,-0.05 v_s^2, 0)$),
for current parallel an perpendicular to $ \vect v_s$, respectively. For the
thin blue curve we assumed 
(as in Fig.~\ref{figRotJTheory1}) that the damping constants are independent of
$t$ while for the
thick green curve a weak temperature
dependence of the damping constant $\alpha$, $\vect \nabla \alpha=0.035 \,\vect
\nabla t$, was assumed. This parameter has
been chosen to reflect the experimental observation, see Fig.~\ref{figExp}.  For
even stronger currents (not measured experimentally and not shown in the figure)
 the size of the torque drops again
 and a finite rotation angle is obtained for $1.57\lesssim v_s \lesssim 2.53$ in the
parallel configuration with the temperature dependent damping constant. 
\label{figRotJTheory2}}
\end{figure}

\begin{figure}[t]
\begin{center}
\includegraphics[width=\linewidth]{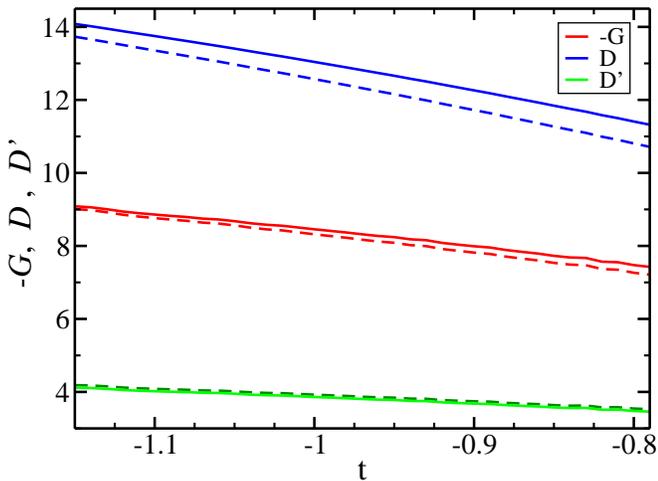}
\end{center}
\caption{Change of $\mathcal{G}, \mathcal{D}$ and $\mathcal{D'}$ defined in Eq.~(\ref{drift1}) with temperature $t$. 
The applied magnetic field is $h/ \sqrt{2} \, (0, 0, 1)$. Dashed lines are for $h=0.9$
and continuous lines for $h=1.1$. 
\label{parameters}}
\end{figure}

Motivated by existing experimental data (discussed below in Sec.~\ref{experiments}) 
we study the case of a temperature gradient, $\lambda=t$, based on the following assumptions.
First, we assume that all damping constants are temperature independent (this assumption is relaxed later). 
Second,
we need also a theory for the temperature dependence of the pinning force. Here
we use the experimental observation~\cite{Schulz12} that the critical current is
almost temperature independent at least for a certain range of temperatures.
Within our theory, Eqs.~(\ref{drift1}) and (\ref{pinningF}), this implies that
all temperature dependence of $\vect F_{\rm pin}$ (i.e., the dependence on the parameter $t$ in Eq. (\ref{tF})) arises from the temperature dependence of the
magnetization $M$ which we calculate from the Ginzburg-Landau theory (\ref{tF}).
From the Ginzburg-Landau theory, we obtain also the temperature dependence of the other parameters, see 
Fig.~\ref{parameters}.

 In
Fig.~\ref{figRotJTheory1}  we show a typical result (for temperature-independent dissipation constants) for the rotation angle and
angular velocity of a skyrmion domain as a function of $\vect v_s$ in the
presence of a temperature gradient.
 For a temperature gradient perpendicular to
the current (lower panel of Fig.~\ref{figRotJTheory1}), the rotation angle increases after the initial jump. For the
gradient parallel to the current, however,  we
obtain that 
the rotation angle {\em drops} after the initial jump (upper panel). For larger values of
$\vect v_s$ the angle rises again until it reaches its maximal value of
$\unit{15}{\degree}$.
This qualitative shape of the curve appears to be rather independent of the
precise values of the various parameters {\em if} we assume that all damping parameters are temperature independent.

 In Fig.~\ref{figRotJTheory2} we plot the rotation angle for small current densities taking an extra effect into account which is present in the experiments described in Ref.~[\onlinecite{Jonietz10}]: as the temperature
gradients are induced by the currents, they grow quadratically with $v_s$. This does not give rise to any qualitative changes. The thin blue curve Fig.~\ref{figRotJTheory2} thereby reflects the same physics as the corresponding curve in Fig.~\ref{figRotJTheory1} (note the different scale on the $x$ axis). 
The thick green curve of Fig.~\ref{figRotJTheory2} shows that one can, however, obtain qualitative different results (an increase rather than a reduction of the rotation angle after the initial jump for $T$ gradients parallel to the current, upper panel)
by including a small temperature dependence of the Gilbert damping $\alpha$. As we will discuss in 
 Sec.~\ref{experiments}, this can reproduce qualitatively the experimentally observed behavior.

\subsubsection{Dependence on orientation of gradients}

 Fig.~\ref{figRotJTheory1} shows that the rotational torques on the system depend
strongly on the relative orientation of gradient and current. More importantly,
one probes different physical mechanism for gradients parallel or perpendicular
to the current. This effect was already discussed in the introduction, see
Fig.~\ref{fig1}, where, however, only the simple case of a static domain without
pinning was described. In reality, the situation is more complex. All
directional information is encoded in the function $\V(v_s)$ which can be
obtained by first solving Eq.~\eqref{drift1} to obtain
$\vect v_d$ and then comparing Eqs.~\eqref{TTT} and \eqref{compactTTT}.
Unfortunately, a rather large number of unknown parameters (most importantly,
the pinning forces and their dependence on $\lambda$) enters the description.
Therefore we will discuss in the following only a few limiting cases.

A drastically simplified picture occurs in regimes when only two forces 
dominate in Eq.~\eqref{drift1}. For example, close to the
pinning transition, the Magnus force is of the same order as the pinning force
while the two dissipative forces are typically much smaller.
In this case  one can use Eq.~\eqref{drift1} to show that $\hat{ \vect v}_d$ becomes proportional
to $\hat{\vect B} \times (\vect v_s - \vect v_d)$.
Thus, for an $\lambda$-independent $\vect v_s$, both the reactive
rotational coupling vector and the rotational pinning vector become proportional
to $\vect \nabla \lambda \cdot (\vect v_s-\vect v_d)$ (here we neglect a 
possible $\lambda$-dependence of $v_s$). Therefore the {\em ratio} of the component 
of $\V$ parallel ($\V^\|$) and perpendicular ($\V^\perp$) to $v_s$ depends only on the 
direction in which the skyrmion lattice drifts. 
\begin{equation}\label{vv}
\frac{\V^\|}{\V^\perp}
\approx  \frac{(\vect v_s-\vect
v_d)^\parallel}{(\vect v_s-\vect v_d)^\perp}=- \frac{v_d^{\perp}}{v_d^{\parallel}}
\end{equation}
The ratio $\frac{\V^\|}{\V^\perp}$ can be obtained experimentally by measuring the rotation angle or the 
angular velocity for
$\vect \nabla \lambda$ parallel and perpendicular to the current, from which one can obtain directly $\frac{\V^\|}{\V^\perp}$ using Eqs.~(\ref{phi}) and (\ref{omega}). For small angles, $\arcsin x \approx x$, for example, one obtains 
$\frac{\V^\|}{\V^\perp}$ directly from the ratio of the two rotation angles. A different, but probably more precise way to determine this ratio is to find experimentally the  ``magic angle'' $\phi_m$ of gradient vs. current, where all rotations vanish,
$\vect \nabla \lambda \cdot \V=0$. In this case one obtains 
\begin{equation}
\frac{\V^\|}{\V^\perp}=\frac{1}{\tan{\phi_m}}
\end{equation}
This should allow for a quantitative determination of  $v_d^{\perp}\, /\, v_d^{\parallel}$. As $v_d^\parallel$ can be measured independently using emergent electric fields generated by the motion of skyrmions \cite{Schulz12}, one can obtain   the complete information on the drift motion by combining both experiments. It is also instructive to compare skyrmions and vortices in a superconductor. Vortices and skyrmions follow essentially the same equation of motions, Eq.~(\ref{drift1}). The relevant parameters (and therefore also the pinning physics) are, however, rather different. For vortices in conventional superconductors\cite{Blatter94,Kopnin02} the dissipation is very large $\mathcal D \alpha \gg \mathcal G$. Therefore, vortices drift -- up to small corrections -- predominantly {\em perpendicular} to the current while for magnetic skyrmions we expect that at least not too close to the depinning transition, the motion is dominantly parallel to the current.

In the limit where the pinning forces can be neglected, i.e., $v_s \gg v_{\mathrm{pin}}$, to linear order in $\tilde{\beta}$ and $\tilde{\alpha}$ the vector $\V$
is given by
\begin{align}
\V &=- \frac{A}{4 \pi \chi}  \left( \hat{\vect B} \times {\vect v}_s
\right) \left( (\tilde{\beta}-\tilde{\alpha}) \frac{\partial \mathcal{G}}{\partial \lambda} 
\frac{\mathcal{D}}{\mathcal {G}} + \frac{\partial \mathcal{D} 
(\tilde{\beta}-\tilde{\alpha})}{\partial \lambda} 
\right)\\
 &=- \frac{A}{4 \pi \chi}  \left( \hat{\vect B} \times {\vect v}_s
 \right) \frac{1}{\mathcal{G}}
\frac{\partial}{\partial \lambda} \left(\mathcal{D} \mathcal{G}
(\tilde{\beta}-\tilde{\alpha})  \right)
\end{align}
Here we also neglected a possible $\lambda$-dependence of $v_s$.
In this limit the rotation can be induced primarily by gradients perpendicular to $\vect v_s$ reflecting
that the motion of skyrmions is mainly parallel to the current, see Eq.~(\ref{vv}) and Eq.~(\ref{vd}). 
This is also consistent with the behavior shown in Fig.~\ref{figRotJTheory1} where we used a two-times smaller gradient for the perpendicular configuration and obtained nevertheless an onset of the rotational motion for values of $v_s$ much smaller than in the parallel configuration.
Note that in a Galilean invariant system, $\tilde \alpha=\tilde \beta$, no torques can be expected.


\subsection{Experimental situation}
\label{experiments}

Our study is directly motivated by recent neutron scattering experiments in the skyrmion lattice phase of MnSi \cite{Jonietz10}. In the presence of a sufficiently large current, a rotation of the magnetic diffraction pattern by a finite angle was observed when simultaneously a temperature gradient was present (only temperature gradients parallel to the current have been studied). The rotation angle could be reversed by reversing either the direction of the current, the direction of the magnetic field or the direction of the temperature gradient. This clearly showed that rotational torques in the experiment were driven by the interplay of gradients and currents as studied in this paper.
 
\begin{figure}[t]
\begin{center}
\includegraphics[width=\linewidth]{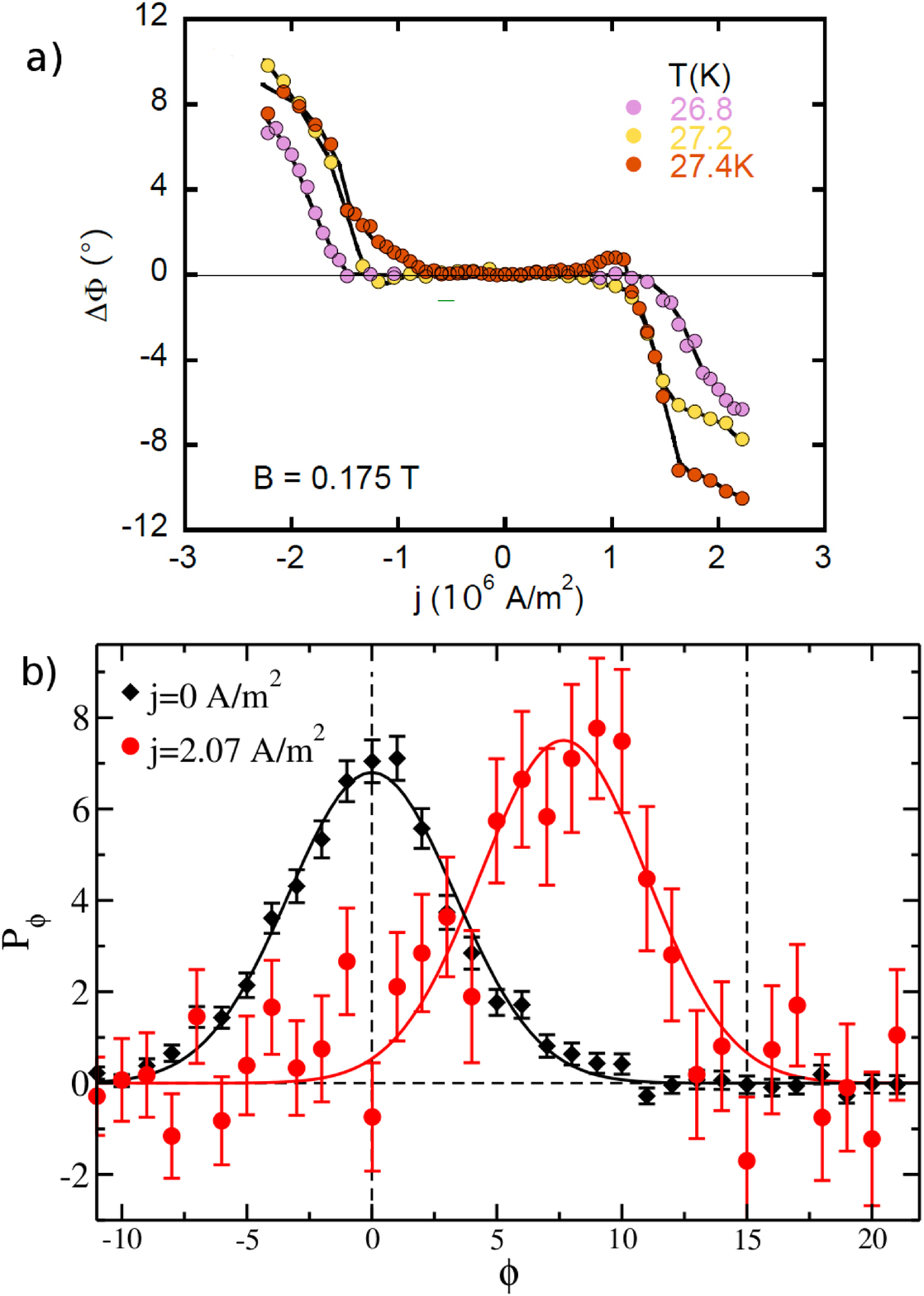}
\end{center}
\caption{a) Average rotation angle $\Delta \phi$ (in units of $\unit{1}{\degree}$) of the skyrmion lattice in MnSi measured by
neutron scattering in the presence of an electric current and a temperature
gradient parallel to the current. The figure is taken from
Ref.~[\onlinecite{Jonietz10}] where further details on the experimental setup can
be found. b) Angular distribution $P_\phi$ of the intensity  normalized to $1$
for currents of strength $j=0$ (black diamonds) and $j\approx -2.07 \cdot 10^6\,$A/m$^2$
for $T=27.4\,$K (red circles). The lines are Gaussian fits serving as a guide to the eye. The
distribution of angles
extents up the maximally possible rotation angle of $\unit{15}{\degree}$ which
suggests that some of the larger domains are rotating with finite angular
velocity for this parameter range.} 
\label{figExp}
\end{figure}

In Fig.~\ref{figExp}a we reproduce Fig.~3~(A) of Ref.~[\onlinecite{Jonietz10}], which shows the average rotation angle (defined as the maximum of the azimuthal distribution of the scattering intensity) as a function of current density. Above a critical current, $j >j_c$, the rotation sets in. The rotation angle initially increases abruptly, followed by a slower increase for larger current densities. When comparing these results with our theory one has to take into account that the temperature gradient in the experiment was not independent of the strength of the applied electrical current density as it originated in the resistive heating in the sample. Therefore the temperature gradient was growing with $j^2$ (i.e. the heating rate due to the electric current). 
This was taken into account in Fig.~\ref{figRotJTheory2} as discussed above. 
For a full quantitative comparison of theory and experiment, it would be desirable to have data, where the applied current as well as both the strength and the direction of the gradients are changed independently. As such data is presently not available, we restrict ourselves to a few more qualitative observations.

In our theory we expect a jump of the rotation angle at $j_c$, which depends on the domain size.  This appears to be consistent with the steep increase of the rotation angle as observed experimentally at $j_c$, especially when taking into account the experimental results are subject to a distribution of domain sizes.

Interestingly, the experimentally observed increase of the rotation angle after its initial jump is apparently {\em not} consistent with the predictions from the extended Landau-Lifshitz-Gilbert equation shown in Eq.~(\ref{LLG}) {\em if} we assume  $\alpha,\alpha',\beta,\beta'$ are independent of temperature. As shown in Fig.~\ref{figRotJTheory2}, we can, however, describe the experimentally observed behavior if we assume a weak temperature dependence of the Gilbert damping.

An important question concerns, whether the existing experiments already include evidence of
 some larger domains that rotate continuously. Fig.~\ref{figExp}a shows that for the largest
 currents {\em average} rotation angles of up to $\unit{10}{\degree}$ have been obtained. 
As this is rather close to the maximally possible value of $\unit{15}{\degree}$ for static domains,
 this suggests that continuously rotating domains are either already present in the system or  may be reached 
by using slightly larger currents or temperature gradients.

We have therefore investigated the angular distribution of the scattering pattern using the same set of experimental data analyzed in Ref.~[\onlinecite{Jonietz10}] (technical details of the experimental setup are reported in this paper). In Fig.~\ref{figExp}b we show the azimuthal intensity distribution with and without applied current. Already for zero current a substantial broadening of the intensity distribution is observed. The origin of this broadening are demagnetization effects which lead to small variations of the orientation of the local magnetic fields in the sample tracked closely by the skyrmions. It has been shown \cite{Adams2011} that this effect can be avoided in thin samples when illuminating only the central part of this sample. 
For the existing data this implies that a quantitative analysis of $P_\phi$ is not possible. 
We observed that the measured experimental distribution of angles extents up just to $\unit{15}{\degree}$.
Therefore, from the present data we can neither claim nor exclude that continuously rotating domains already exist
for this set of data but slightly larger current densities or gradients should be sufficient
to create those.

\section{Conclusions}

The magnetic skyrmion lattices, first observed in MnSi, have by now been
observed in a wide range of cubic, chiral materials including
insulators~\cite{Seki12,Adams12}, doped semiconductors~\cite{muen09} and good 
metals~\cite{mueh09,pfleiderer:JPCM2010}. This is expected
from theory: in any material with B20 symmetry, which would be ferromagnetic in
the absence of spin-orbit coupling, weak Dzyaloshinskii Moriya interaction
induce skyrmion lattices in a small magnetic field. While in bulk they are only
stabilized in a small temperature window by thermal fluctuations
close to the critical temperature, they are much more stable in thin
films~\cite{yu:2010, yu:2011}. 

From the viewpoint of spintronics, such skyrmions are ideal model systems to investigate the coupling of electric-, thermal- or spin currents to magnetic textures: (i) the coupling by Berry phases to the quantized winding number provides a universal mechanism to create efficiently Magnus forces, (ii) skyrmion lattice can be manipulated by extremely small forces induced by ultrasmall currents \cite{Jonietz10,Schulz12}, (iii) the small currents imply that also new types of experiments (e.g., neutron scattering on bulk samples) are possible. 

We think that the investigation of the rotational dynamics of skyrmion domains
provides a very useful method to learn in more detail which forces affect the dynamics of the magnetic texture. As we have shown, the rotational torques can be controlled by both the strength and the direction of field- or temperature gradients in combination with electric currents. They react very sensitively
not only on the relative strength of the various forces but also on how the forces depend on temperature and field. 

While some aspects of the theory, e.g. the dependence on the strength of the gradients, can be worked out in detail, many other questions remain open. An important question is, for example, to identify the leading 
damping mechanisms and their dependence on temperature and field. Also an understanding of the interplay of pinning physics, damping and the motion of magnetic textures is required to control spin torque effects.
Here future rotation experiments are expected to give valuable information. Furthermore, it will be interesting to study the pinning physics in detail and to learn to what extent 
skyrmions and vortices in superconductors behave differently.

One way to observe the rotation of the skyrmion lattice is to investigate the angular distribution of the neutron 
scattering pattern as discussed in Sec.~\ref{phiandomega}. This does, however, only provide indirect evidence on the expected continuous
rotation of the skyrmion lattice. Therefore it would be interesting to observe the continuous rotation more directly.
For example, one can use that time-dependent Berry phases arising from moving skyrmions induce ``emergent''
electrical fields which can be directly measured \cite{Schulz12} in a Hall 
experiment. Here it would be interesting to observe higher harmonics in the 
signal which are expected to appear close to the threshold where continuous rotations set in, see Fig.~\ref{figAngle}.

In future, it might also be interesting to use instead of electrical current other methods, e.g. pure spin currents or thermal currents, to manipulate skyrmion lattices (e.g. in insulators). We expect that also in such systems the investigation of rotational motion driven by gradients will give useful insight in the control of magnetism beyond thermal equilibrium.


\acknowledgments
We gratefully acknowledge discussions with M. Halder, M. Mochizuki and T. Nattermann. We also acknowledge financial support of the German Science Foundation (DFG) through SFB 608, TRR80 and FOR960, as well as the European Research Council through ERC-AdG (291079). KE wishes to thank the Deutsche Telekom Stiftung and the Bonn Cologne Graduate School.


\bibliography{spintorque}

\begin{thebibliography}{35}
\bibitem[{foo({\natexlab{a}})}]{footnoteSpinCurrent}
\bibinfo{note}{The spin current $j_s^{j i}$ is a tensor describing the current
  of the spin component $j$ flowing in the direction $i$. To describe spin
  torques, the projection of $j_s^{j i}$ on the local direction of the
  magnetization is needed, $j_s^i= \hat{\vect \Omega}_j j_s^{j i}$ and $\vect
  v_s \approx \vect j_s/|\vect M|$.}
\bibitem[{foo({\natexlab{b}})}]{footnote1}
\bibinfo{note}{In previous publications \cite{everschor11} we have defined
  $\vect G$ and $\vect D$ without the factors $M$ which need to be included
  here for the discussion of gradients.}
\end{thebibliography}
\bibliographystyle{apsrev}

\end{document}